\documentclass[pre-print,3p,times,fleqn]{elsarticle}
\usepackage{amsmath}
\usepackage{amssymb}
\usepackage{psfrag}
\usepackage{color}
\usepackage{stfloats}
\usepackage{fixltx2e}
\usepackage{mathrsfs}
\usepackage{tabularx}
\usepackage{booktabs}
\usepackage{colortbl}
\usepackage{rotating}
\usepackage{boldline}
\usepackage{bm}
\usepackage{changes}
\usepackage{url}

\DeclareMathOperator{\Tr}{Tr}

\usepackage{subfigure}
\usepackage{lineno}

\usepackage{soul}
\journal{arXiv}
\begin{document}
\renewcommand{\topfraction}{0.98}	
\renewcommand{\bottomfraction}{0.98}
\setcounter{topnumber}{3}
\setcounter{bottomnumber}{3}
\setcounter{totalnumber}{4}         
\setcounter{dbltopnumber}{4}        
\renewcommand{\dbltopfraction}{0.98}	
\renewcommand{\textfraction}{0.05}	
\renewcommand{\floatpagefraction}{0.5}	
\renewcommand{\dblfloatpagefraction}{0.5}	
\newcommand{\beq}{\begin{equation}}
\newcommand{\eeq}{\end{equation}}
\newcommand{\divg}{\mbox{\rm{div}}\,}
\newcommand{\Divg}{\mbox{\rm{Div}}\,}
\newcommand{\D}  {\displaystyle}
\newcommand{\DS} {\displaystyle}
\newcommand{\RM}[1]{\textit{\MakeUppercase{\romannumeral #1{}}}}
\newtheorem{remark}{\bf{{Remark}}}
\def\sca   #1{\mbox{\rm{#1}}{}}
\def\mat   #1{\mbox{\bf #1}{}}
\def\vec   #1{\mbox{\boldmath $#1$}{}}
\def\scas  #1{\mbox{{\scriptsize{${\rm{#1}}$}}}{}}
\def\scaf  #1{\mbox{{\tiny{${\rm{#1}}$}}}{}}
\def\vecs  #1{\mbox{\boldmath{\scriptsize{$#1$}}}{}}
\def\tens  #1{\mbox{\boldmath{\scriptsize{$#1$}}}{}}
\def\tenf  #1{\mbox{{\sffamily{\bfseries {#1}}}}}
\def\ten   #1{\mbox{\boldmath $#1$}{}}
\def\Ass  {\overset{\hspace*{0.4cm} n_{\scas{el}}}
          {\underset{\scaf{c},\scaf{d}=1}{\msf{A}}}}
\def\ltr   #1{\mbox{\sf{#1}}}
\def\bltr  #1{\mbox{\sffamily{\bfseries{{#1}}}}}
\sloppy
\begin{frontmatter}
\title{\Large WarpPINN: Cine-MR image registration with physics-informed neural networks.}


\author[01]{Pablo Arratia L\'opez}
\author[03,08]{Hern\'an Mella}
\author[08,04]{Sergio Uribe}
\author[05,06]{Daniel E. Hurtado}
\author[08,06,07]{Francisco Sahli Costabal}
\ead{fsc@ing.puc.cl, corresponding author}

\address[01]{Department of Mathematical Sciences, University of Bath, Bath, UK}

\address[03]{School of Electrical Engineering, Pontificia Universidad Católica de Valparaíso, Valparaíso, Chile}

\address[08]{Millennium Institute for Intelligent Healthcare Engineering, iHEALTH}
\address[04]{Biomedical Imaging Center, Pontificia Universidad Católica de Chile, Santiago, Chile}

\address[05]{Department of Structural and Geotechnical Engineering, School of Engineering, Pontificia Universidad Cat\'olica de Chile, Santiago, Chile}

\address[06]{Institute for Biological and Medical Engineering, Schools of Engineering, Medicine and Biological Sciences, Pontificia Universidad Cat\'olica de Chile, Santiago, Chile}

\address[07]{Department of Mechanical and Metallurgical Engineering, School of Engineering, Pontificia Universidad Cat\'olica de Chile, Santiago, Chile}



%
\begin{abstract} %
Heart failure is typically diagnosed with a global function assessment, such as ejection fraction. However, these  metrics have low discriminate power, failing to distinguish different types of this disease. Quantifying local deformations in the form of cardiac strain can provide helpful information, but it remains a challenge. In this work, we introduce WarpPINN, a physics-informed neural network to perform image registration to obtain local metrics of the heart deformation. We apply this method to cine magnetic resonance images to estimate the motion during the cardiac cycle. We inform our neural network of near-incompressibility of cardiac tissue by penalizing the jacobian of the deformation field. The loss function has two components: an intensity-based similarity term between the reference and the warped template images, and a regularizer that represents the hyperelastic behavior of the tissue. The architecture of the neural network allows us to easily compute the strain via automatic differentiation to assess cardiac activity. We use Fourier feature mappings to overcome the spectral bias of neural networks, allowing us to capture discontinuities in the strain field. We test our algorithm on a synthetic example and on a cine-MRI benchmark of 15 healthy volunteers. We outperform current methodologies both landmark tracking and strain estimation. We expect that WarpPINN will enable more precise diagnostics of heart failure based on local deformation information. Source code is available at \url{https://github.com/fsahli/WarpPINN}

\end{abstract}
\begin{keyword}
Cardiac strain; Cardiac mechanics; Image Registration; Physics-Informed Neural Networks
\end{keyword}
\end{frontmatter}


\section{Motivation}\label{intro}

Heart disease is one of the leading causes of death in the world, taking around 17.9 million lives in 2016 according to the World Health Organization \cite{OMS}. Alterations in the mechanical function of the heart may trigger complex diseases in the body. These changes may be caused by irregularities in the biophysical, electrical and/or cellular processes involved in the heart's function. For instance, pulmonary hypertension is characterized by abnormal contractions of the right ventricle \cite{von_Siebenthal_Aubert_Mitsakis_Yerly_Prior_Nicod_2016}. Thus, understanding cardiac motion to detect subtle changes with precision and improve diagnosis and prognosis of heart disease is a task of major interest.

Currently, there are multiple metrics to assess the mechanical performance of the heart. In particular, global function of the left ventricle is widely used to assess heart failure. However, the information provided by these simplified quantities is not enough to characterize heart disease. The most commonly used metric is the left ventricle ejection fraction (LVEF), which measures the portion of blood ejected relative to the end diastolic volume. Thus, LVEF is an indicator of the pumping function of the heart and may help to diagnose and track heart failure. It is fairly common that a patient presents normal LVEF percentage, between 50\% and 70\%, but still presents heart failure. This phenomenon is known as heart failure with preserved ejection fraction (HFpEF), also referred to as diastolic heart failure. This condition may arise if cardiac tissue becomes stiffer, which may reduce the amount of blood entering the ventricle, and the amount of ejected blood might also be reduced while ejection fraction remains in acceptable level. This condition highlights the importance of developing more advanced metrics for cardiac activity based on other quantities. \cite{Pfeffer_Shah_Borlaug_2019}

In order to study and measure the cardiac motion, several imaging techniques have been developed, such as echocardiography, positron emission tomography (PET), computed tomography (CT), or magnetic resonance imaging (MRI) \cite{Wayne_Alexander_Schlant_Fuster_ORourke_Roberts_Sonnenblick_Russell_2000}. Echocardiography is the most common method to measure LVEF but currently, MRI is considered the gold standard method for this task \cite{Marcu_Beek_van_Rossum_2006}. MRI and cine steady-state free precession (SSFP) MRI, in particular, are one of the most widely used non-invasive techniques employed to qualitatively and quantitatively assess heart functionality mainly because of their benefits such as high soft-tissue contrast, high reproducibility, and good spatial and temporal resolution \cite{BISTOQUET_OSHINSKI_SKRINJAR_2008}. MRI can be used to measure chamber or ventricular volumes, myocardial mass, myocardial viability, ejection fraction, etc. As opposed to other MRI techniques, cine SSFP MRI does not contain direct information about cardiac deformation. For instance, tagged \cite{Axel2005}, phase-contrast velocity mapping (tissue phase mapping) \cite{pelc_evaluation_1994} and DENSE MRI \cite{Aletras1999,Kim2004} are three imaging approaches that encode the tissue motion into the image, allowing a more direct estimation of ventricular deformation. However, in contrast to cine SSFP MRI, these techniques are not widely used in clinical routines because their elevated scan times. For these reasons, there have been many efforts to accurately measure cardiac motion from cine SSFP MR images \cite{Bello_2019, BISTOQUET_OSHINSKI_SKRINJAR_2008, DBLP:journals/corr/abs-1808-08578}. 

Cardiac strain has gained special interest in the past decades as an advanced measure of cardiac function due to its good reproducibility \cite{Bucius2020} and sensitivity to subtle changes in cardiac contractility. This quantity describes the relative displacement of cardiac tissue through the whole contraction \cite{Scatteia2017} and has been extensively studied in the context of several diseases. Cardiac strain has demonstrated being a powerful tool for the assessment and diagnosis of conditions such as heart failure, cardiomyopathies, dyssynchrony, abnormal pressures, valve lesions, ventricular arrhythmias, and dysfunction in hemodialysis patients among others \cite{Chitiboi2017,kalam2014prognostic,ChiuA90}. Usually, cardiac strain is reported using three regional metrics named Circumferential, Radial, and Longitudinal strains that represent tissue deformations in different directions. We are interested in regional strains, corresponding to the average of the strains throughout time defined over a selected set of AHA segments \cite{Selvadurai2018} and which provide with relevant information about the cardiac function. The strong evidence of the prognostic value of strain-based metrics versus more simplistic metrics such as LVEF motivates obtaining cardiac strains from cine SSFP MRI. It has the potential to elucidate novel biomarkers for cardiac disease, with this information coming at no extra cost, as cine SSFP MRI is performed routinely to assess cardiac function.

To solve this problem, we propose using physics-informed neural networks (PINNs) \cite{Raissi_Perdikaris_Karniadakis_2019,sahli2020physics}. This method excels when there is incomplete data but some physical model of the observed phenomenon is available. In general, a fully-connected neural network is trained to satisfy a data fidelity term and a physics term, which encodes the model in some differential operator. Here, we propose WarpPINN, where we will approximate the deformation mapping from end-diastole to any point in the cardiac cycle with a neural network and train it to solve an image registration problem and to satisfy the near-incompressibility of cardiac tissue. 

This paper is organized as follows: section 2 is devoted to explain the methodology proposed to solve the problem, introducing the main concepts such as image registration, quasi-incompressibility of the myocardium, physics-informed neural networks, WarpPINN and Fourier feature mappings. In section 3, an explicit and fully-incompressible deformation field is constructed and applied on a bidimensional image to perform a controlled synthetic experiment. In section 4 we apply WarpPINN on real cine-MRI data to obtain strain metrics from the predicted deformation field and present the main results. Finally, in section 5 we address the main advantages and drawbacks of our method and discuss what is left for future research.

\section{Methods}

\subsection{Image registration}

In order to quantify the cardiac motion and compute cardiac strains from cine SSFP MRI, we solve an image registration problem. The goal is to match two or more images displaying the same object measured at different times by finding the most suitable alignment or transformation to achieve point-wise spatial correspondence \cite{10.1145/146370.146374, Modersitzki_2003,BarnafiEtal2018}. Typically, there are two images, one is called reference, and the second is called template. The template image $T:\Omega_1\subseteq\mathbb{R}^n\to \mathbb{R}$ (for images we take $n=2,3$) has to be composed with an unknown displacement field $\vec{\varphi}:\Omega_0\to\Omega_1$ such that  $T\circ \vec{\varphi}$ matches with the reference image $R:\Omega_0\subseteq\mathbb{R}^n\to \mathbb{R}$. In other words, the registration task may be established as follows: given the reference image $R$ and the template image $T$, find the transformation $\vec{\varphi}$ such that $T\circ\vec{\varphi}$ is similar to $R$ in a suitable way. From a mathematical point of view, the registration task may be stated as an optimization problem where we seek for a minimizer $\varphi$ of the following expression:
\begin{equation}\label{eq:R - T phi}\|R-T\circ \vec{\varphi}\|^p_{p} + \mu \mathcal{R}(\vec{\varphi}).\end{equation}

In equation \eqref{eq:R - T phi} the notion of similarity is given by a norm $\|\cdot\|_p$ for some suitable normed space $p$. This problem is ill-posed and then additional information about the displacements $\vec{\varphi}$ is encoded in the regularization term $\mu \mathcal{R}(\vec{\varphi})$ to obtain a meaningful solution. 

The transformation $\vec{\varphi}$ may be rigid or non-rigid. In rigid transformations, the deformation is global, meaning that the entire image is transformed according to a specific criteria, for example rotations, translations or shearing deformations. Non-rigid transformations can warp the template image locally with, for instance, radial basis functions, physical continuum models or large deformation models \cite{10.5555/1050969, Modersitzki_2003}. 

In our case, we are interested in tracking the tissue motion of the heart through the cardiac cycle. We consider the reference image as the snapshot of cine SSFP MRI at end-diastole, while the template image is any image of the sequence at some later moment. Then, with a multi-temporal registration and non-rigid transformations we can determine the deformation field between the initial frame and the rest of the frames, where the non-rigid transformation will be given by our neural network, WarpPINN, which, once trained, will directly compute the strains. 

\subsection{Incompressibility of the myocardial tissue}

The incompressibility of cardiac tissue has recently become a debated topic in the literature. Early studies in canines showed that it is not exactly incompressible, as the volume in the capilaries can be reduced, changing the total volume by 2-4\% during contraction \cite{Rodriguez_Ennis_Wen_2006, Yin_Chan_Judd_1996}. More recent studies based on cine MRI report a decrease in volume from diastole to systole of around 12\% in healthy subjects and volume preservation, on average, for some diseased patients \cite{ryu2021systolic,kumar2021cardiac}. Other studies based on cine MRI suggest an increase in myocardial volume from diastole to systole between 10 and 20\% \cite{orlowska2017systolic}. One limitation of these studies is the lack of spatial resolution in the short axis (8 mm), which may induce a bias in the results. Studies using fiduciary markers in ovine models also show mixed results: one study shows moderate local changes in volume from diastole to systole, between -2\% and +3\% on average \cite{tsamis2012kinematics}, while other study on ovine as well showed a 15\% reduction in volume \cite{avazmohammadi2020vivo}. Overall, we assume that while the myocardium is not exactly incompressible is also not a highly compressible material, so local volume changes should remain controlled. We will use this fact to inform our neural network and improve our strain estimations by penalizing large volumetric deformations.   

\subsection{WarpPINN: physics-informed neural networks for image registration}

\begin{figure}[ht]
\centering
\includegraphics[width=\textwidth]{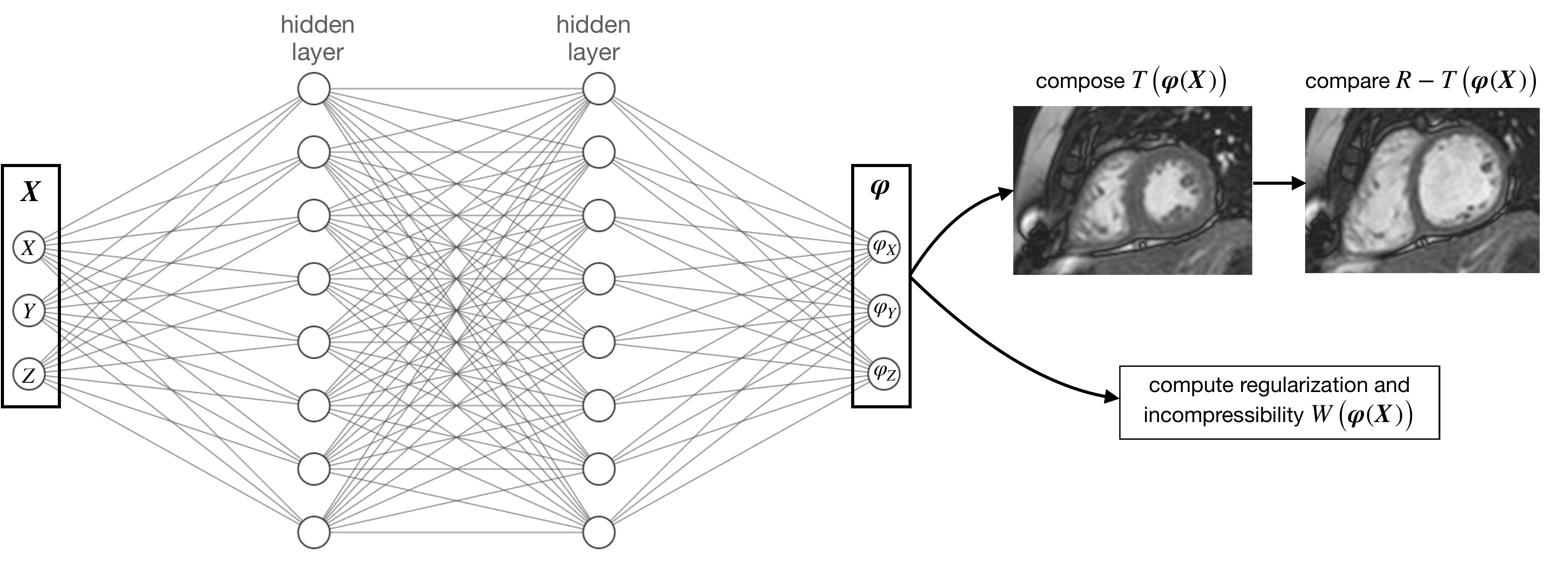}
\caption{A neural network predicts the displacement field $\vec{\varphi}$ as a function of the position in the image $\vec{X}$. This network is trained to find the displacement field that transforms the end diastolic image to end systolic image, while respecting the incompressibility of cardiac tissue.}
\label{fig:PINN}
\end{figure}

Nowadays, deep learning constitutes one of the most important branches of machine learning. The explosive growth of this field is partly because of the huge impact that has had in a wide range of areas such as image recognition, medical imaging, natural language processing, automatic speech recognition, etc. \cite{poplin2018prediction}, and also due to the huge amount of available data we count for different phenomena. However, this is not the situation for physical or biological systems where data acquisition is, more often than not, expensive and hard to obtain. The lack of large datasets makes it difficult to employ standard machine learning techniques. In spite of this, in many systems we have knowledge of the physics of the process, which is typically embedded in the form partial differential equations and can be incorporated into the neural network training process. This novel approach is referred to as physics-informed neural networks (PINNs) and they constitute a new and increasingly popular sub-field of scientific machine learning \cite{Raissi_Perdikaris_Karniadakis_2019, Berg_2018, Dissanayake_Phan-Thien_1994}. In these neural networks, we look for a function $u(\vec{x})$ satisfying both the available data and the corresponding partial differential equation. Let $\mathcal{N}$ be a differential operator acting on $u$ (for instance the Laplacian $\nabla \cdot \nabla$) and let $\mathcal{B}$ be a boundary differential operator (for example Dirichlet, Neumann or Robin boundary conditions), such that the prior knowledge that we have of $u$ is given by the next equations  
\[\mathcal{N}[u](\vec{x}) = 0, \forall \vec{x} \in \Omega, \quad \mathcal{B}[u](\vec{x}) = g(\vec{x}), \forall \vec{x} \in \partial\Omega,\]
where $\Omega$ is the domain and $\partial\Omega$ its boundary. We aim to approximate $u$ with a neural network $u(\vec{x}; \vec{\theta})$ parameterized by the weights $\vec{\theta}$. To achieve this goal, the input of the network is a spatial point $\vec{x}$ of the domain (it may include the time variable for dynamical systems) and the output is the approximated value of $u$ at this point. See figure \ref{fig:PINN}. In general, data of the quantity of interest $\hat{u}_i$ will be available only at a small number of locations $\{\vec{x}^d_i\}_{i=1}^{N_d}\subset\Omega$, and we will have information of the boundary conditions $\{\hat{g}_j\}_{j=1}^{N_b}$ at $\{\vec{x}^b_j\}_{j=1}^{N_b}\subset \Omega$. Finally, we count with randomly sampled collocation points $\{\vec{x}^c_k\}_{k=1}^{N_c}\subset \Omega$, where the neural network is trained to satisfy the PDE $\mathcal{N}[u](\vec{x}) = 0$. These points constitute the training data and then we want to minimize over $\vec{\theta}$ a loss function of the form:
\[
\mathcal{L}(\vec{\theta}) = \dfrac{1}{N_d}\displaystyle\sum_i^{N_d} (u(\vec{x}^d_i;\vec{\theta}) - \hat{u}_i)^2 + \dfrac{1}{N_b}\displaystyle\sum_{j}^{N_b}(\mathcal{B}[u](\vec{x}^b_j;\vec{\theta})-\hat{g}_j)^2 +
\dfrac{1}{N_c}\displaystyle\sum_{k}^{N_c}(\mathcal{N}[u](\vec{x}^c_k;\vec{\theta}))^2, 
\]
where the first term fits the data, the second term fits the boundary condition, and the third term fits the differential equation.

For these applications, automatic differentiation is fundamental not only because it allows to apply gradient-based routines by calculating the derivative of the output with respect to the weights $\vec{\theta}$, but also because it enables the differentiation of the output with respect to the input $\vec{x}$ to compute the derivatives involved in the differential operators. 

In this work, we represent the displacement field with a fully connected neural network $\vec{u}(\vec{X};\vec{\theta})$, such that $\vec{\varphi}(\vec{X};\vec{\theta}) = \vec{X} +  \vec{u}(\vec{X};\vec{\theta})$. We incorporate the physical knowledge by including 2 hyperelastic regularization terms: one for cardiac tissue, which will favor quasi-incompressibility and one for the background. Then, we train the neural network by minimizing the following loss function: 
\begin{equation}\label{eq:loss function 1}
\begin{array}{rcl}
 \mathcal{L}(\vec{\theta}) &=& \|R-T\circ \vec{\varphi}\|_p^p + \mu\mathcal{R}(\vec{\varphi}) \\  &=& \dfrac{1}{N_d} \displaystyle\sum_i^{N_d} (R(\vec{X}^d_i) - T(\vec{\varphi}(\vec{X}^d_i;\vec{\theta})))^p + \mu \left( \frac{1}{N_{\rm inc}}\displaystyle\sum_{l}^{N_{\rm inc}} W(\vec{\varphi}(\vec{X}^{{\rm inc}}_l; \vec{\theta});\lambda_{\rm inc}) + \frac{1}{N_{\rm bg}}\displaystyle\sum_{k}^{N_{\rm bg}} W(\vec{\varphi}(\vec{X}^{{\rm bg}}_k; \vec{\theta});\lambda_{\rm bg})\right),
 \end{array}
\end{equation}
where $\vec{X}^d_i$ represents the position of the i-th pixel in the reference image $R$ of size $N_d$. The points $\{\vec{X}^{\rm inc}_l\}_{l=1}^{N_{\rm inc}} \subset \Omega_{\rm inc}$ and $\{\vec{X}^{\rm bg}_k\}_{k=1}^{N_{\rm bg}} \subset \Omega_{\rm bg}$ in equation \eqref{eq:loss function 1} are collocation points, with $\Omega_{\rm inc}$ being the nearly-incompressible  region where the cardiac tissue is located and $\Omega_{\rm bg}$ being the background of the image. The strain energy function $W$ will act as a regularizer for the displacement predicted by the network. In particular, we choose a neo-Hookean hyperelastic function, with different properties in $\Omega_{\rm inc}$ and $\Omega_{\rm bg}$. In the three dimensional case, this particular function has the form:
\begin{equation}\label{eq:neo hookean}
    W(\vec{\varphi}; \lambda) = \Tr (\vec{C})-3-2\log(J)+\lambda (J-1)^2.
\end{equation} 
Here, we have introduced the determinant of the jacobian of the deformation gradient $J = \det \ten{F}$, with $\ten{F} = d \vec{\varphi}/d \vec{X}$ and the first invariant of the right Cauchy-Green deformation tensor $\ten{C} = \ten{F}^T\ten{F}$. The Jacobian represents changes in volume in the material, where $J = 1$ means volume conservation, i.e., an incompressible behaviour. The first term of the loss function penalizes the difference between the reference image $R$ and the template image $T$ composed with the deformation mapping $\vec{\varphi}$. The term $\lambda (J-1)^2$ penalizes changes in volume, both positive and negative, and is zero when the incompressibility is satisfied. For this particular strain energy, the parameter $\lambda$ will control how strongly the incompressibility is enforced, which will yield to different parameters for the cardiac tissue $\lambda_{\rm inc}$ and the background of the image $\lambda_{\rm bg}$.


\subsubsection{Fourier Feature Mappings}\label{subsec:fourier features}

\begin{figure}[!t]
\centering
\includegraphics[width=\textwidth]{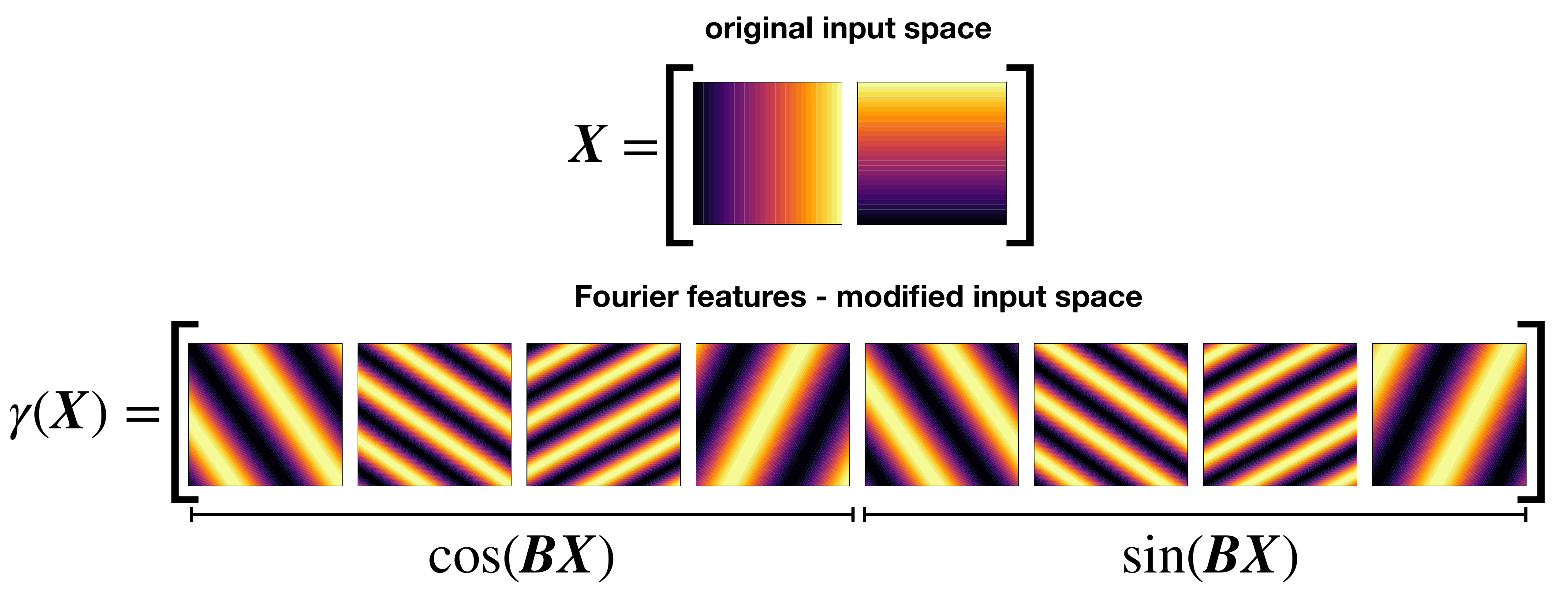}
\caption{Fourier features help overcome the spectral bias of neural networks by introducing modified input space with higher frequencies.}
\label{fig:fourierfeatures}
\end{figure}

A well-documented issue with neural networks is the spectral bias \cite{pmlr-v97-rahaman19a}. During training they tend to learn the low-frequency component of the mapping first, while learning the high-frequency features might take an extremely large amount of training time which finally leads to inaccurate predictions. To overcome this problem, the input of the network is modified with Fourier feature mappings to include higher frequencies \cite{wang2020eigenvector, tancik2020fourier}. Here, we let $d$ be the input dimension, $m\in \mathcal{N}$ and $\sigma>0$ be two hyperparameters, then we define $\ten{B}\in\mathbb{R}^{m\times d}$ where each entry $\ten{B}_{ij}$ is sampled independently from a normal distribution $\mathcal{N}(0,\sigma^2)$. In particular, these are not trainable parameters, and $m$ and $\sigma>0$ indicate the amount of Fourier features to use and the variance of the Gaussian distribution respectively. The Fourier feature mapping $\gamma$ is then defined as
\begin{equation}\label{eq:fourier feature mapping}
\gamma(\vec{X}) = \begin{bmatrix} \cos(\ten{B}\vec{X}) \\ \sin(\ten{B}\vec{X}) \end{bmatrix}.
\end{equation}
Thus, we use a higher dimensional vector $\gamma(\vec{X})\in\mathbb{R}^{2m}$ as input of the network instead of the original vector $\vec{X}\in\mathbb{R}^d$, as shown in Figure \ref{fig:fourierfeatures}). The hyper-parameter $\sigma$ plays an important role: the larger it is, the higher the sampled frequencies, and then sharper features can be predicted. However, one must be careful as very large frequencies may lead to an over-fitting behaviour.  


\section{Numerical examples in 2D}\label{sec:2d}
 
In this section we study a synthetic image registration problem where we construct a deformation field $\vec{\varphi}$ that is incompressible in a region and compressible on the rest of the image. We call WarpPINN our neural network without Fourier features and WarpPINN-FF to the neural network using Fourier features as explained in section \ref{subsec:fourier features}. We study how both implementations approximate the ground-truth deformation field given a reference and template image.
 
\subsection{Image synthesis}
 
A region in the reference image corresponding to a ring of thickness $R_2-R_1$ is deformed into another ring centered at the same point but with thickness $r_2-r_1$ such that the area of both rings is conserved, that is
\begin{equation}\label{eq:same area}
    R_2^2-R_1^2 = r_2^2-r_1^2.
\end{equation}

We propose a radial deformation field of the form
\begin{equation}\label{eq:deformation phi2}
    \vec{\varphi}(X,Y) =  f\left(\sqrt{X^2+Y^2}\right)\begin{pmatrix}
X\\Y
\end{pmatrix}, \quad f(R)=\left\{\begin{array}{ccl} 
\dfrac{r_1}{R_1} & \text{if} & R<R_1,\\
\dfrac{1}{R}\sqrt{R^2-R_1^2+r_1^2}& \text{if} & R_1\leq R \leq R_2,\\
\dfrac{r_2}{R_2} & \text{if} & R>R_2,\\
\end{array}\right.
\end{equation}
where $R = \sqrt{X^2+Y^2}$. The values of $f$ outside the ring makes it a continuous function with discontinuities of the derivative at $R=R_1$ and $R=R_2$. The incompressibility in the ring $R_1^2<X^2+Y^2<R_2^2$ is shown in appendix A. The domain in the reference image is taken as the unit square $\Omega_0=[0,1]\times[0,1]$, then, this radial transformation is done from the point $(0.5,0.5)$ by just considering $\vec{\varphi}(X-0.5,Y-0.5)$. The deformation field and its Jacobian are shown in Figure \ref{fig:jac}, left panel. The intensity of template image is defined as:
\[T(X,Y)=\sin(4\pi X) \cdot \cos(4\pi Y),\]
while the the reference image is simply constructed as 
\[R(X,Y) = T \circ \vec{\varphi} (X,Y).\]

\begin{figure}[h]
\centering
\includegraphics[width=0.8\linewidth]{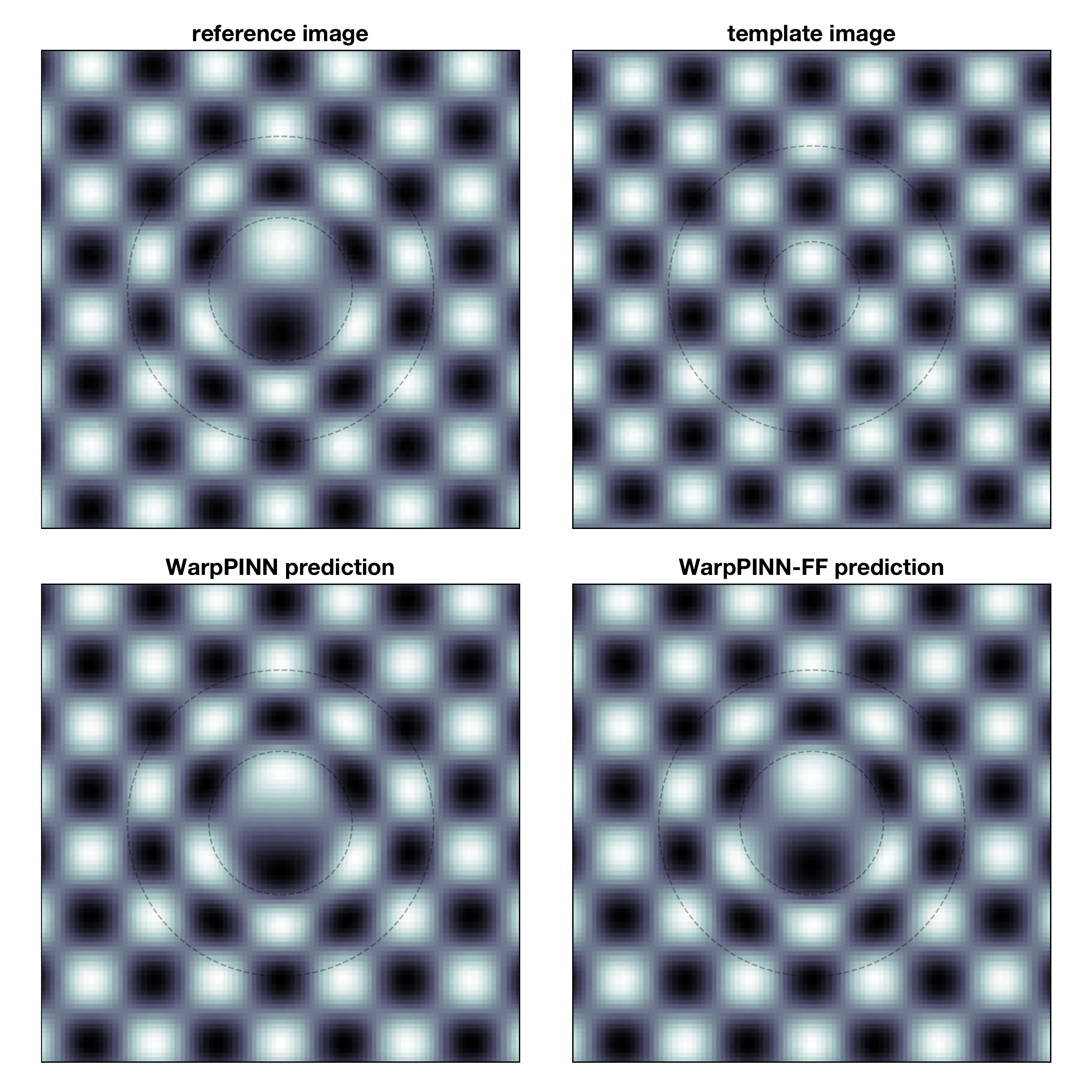}
\caption{Top row: reference and template images respectively. Bottom row: predicted reference image warped by WarpPINN and WarpPINN-FF respectively. The radii in the reference image are $R_1=0.15$, and  $R_2=0.32$, while in the template image are $r_1=0.1$, and $r_2=0.3$.}
\label{fig:Reference and Target ring}
\end{figure}

Reference and template images are shown at the top left and top right in figure \ref{fig:Reference and Target ring} respectively, where we take $R_1=0.15, R_2=0.32, r_1=0.1$, and $r_2=0.3$. This choice satisfies the area conservation in \eqref{eq:same area}. Gaussian noise of 1\% was added to both images to avoid inverse crime.

\subsection{Training the neural network}

In this setting, the neural network approximates a vector field $\vec{u} = (u_1,u_2)$ such that  $\vec{\varphi}(\vec{X};\vec{\theta}) = \vec{X}+\vec{u}(\vec{X};\vec{\theta})$. Hence, it takes two values as input, the coordinates of the point $\vec{X}$, and outputs the two components of $\vec{u}$.

Weights are initialized with Xavier initialization \cite{pmlr-v9-glorot10a} while bias vectors are initialized as 0. As a first step, we train the network to predict close to zero displacement in the entire domain by minimizing the following loss function:
\[ \dfrac{1}{N_d}\displaystyle\sum_{i}^{N_d}\| \vec{u}(\vec{X}^d_i) \|_2^2. \]
This process avoids potential problems that may arise from the random initialization of the neural network, such as displacement fields that produce negative volumes.
Next, we use the loss function defined in \eqref{eq:loss function 1} to train the neural network with $p=2$. In the 2D case, the neo-Hookean introduced in \eqref{eq:neo hookean} takes the form
\[\vec{W} = \Tr(\vec{C})-2-2\log(J)+\lambda(J-1)^2.\]

The architecture considered has dimension two for the input and output layers, and has three hidden layers with 32 neurons each. When using Fourier features, we set  $m=8$ and $\sigma = 10$. To evaluate the similarity term between the reference and the warped image, the forward pass is performed over all the pixel locations and then the template image is deformed using bilinear interpolation of the pixel values at every iteration. For the evaluation of the neo-Hookean 100 collocation points are randomly sampled in the reference image at each iteration. As the incompressible region is assumed known, these points are classified depending on whether they lie or not on the ring to compute the neo-Hookean with the corresponding values $\lambda_{\rm inc}=1000$ if the point is on the ring or $\lambda_{\rm bg}=1$ if the point is on the background. This choice of values for $\lambda_{\rm inc}$ and $\lambda_{\rm bg}$ enforce near-incompressibility on the ring. Finally, the optimiser used is Adam with learning rate $10^{-3}$.

\subsection{Results}

\begin{figure}[t]
\centering
\includegraphics[width=\textwidth]{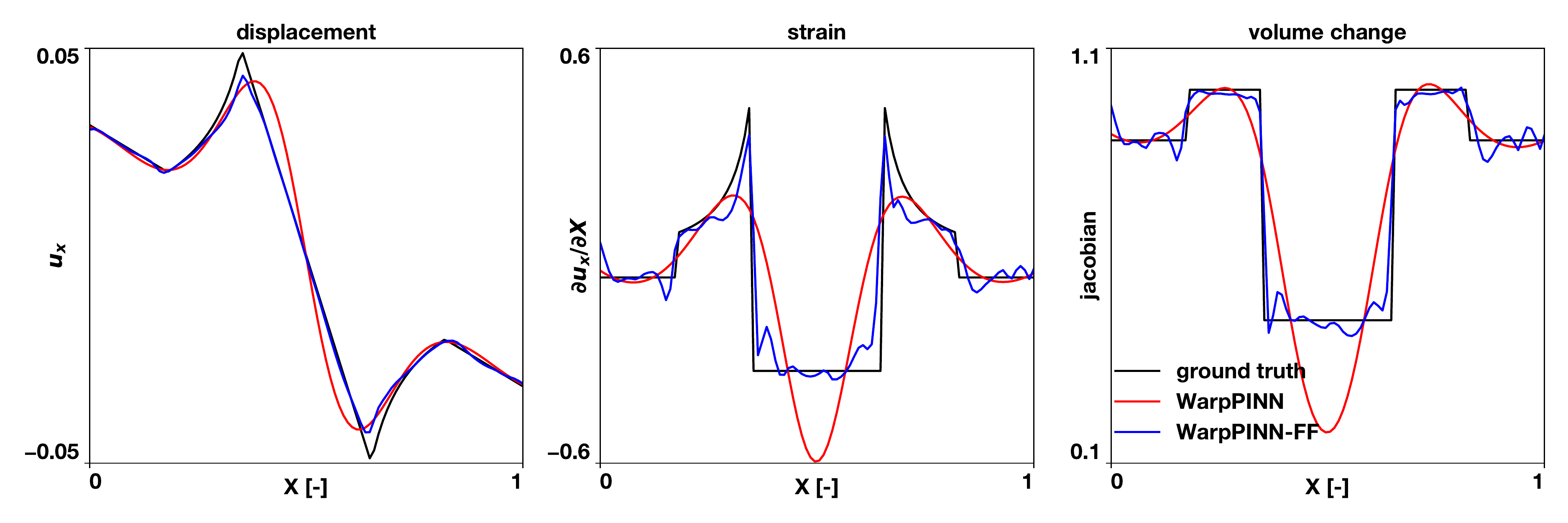}
\caption{Profiles at $y  =0.505$ of $u_x$ (left),  $\partial u_x/\partial X$ (center) and jacobian (right). WarpPINN-FF (blue) captures discontinuities in derivatives better than WarpPINN (red). Discontinuities on reference strain and jacobian occur at the boundaries of the ring.}
\label{fig:jac slice}
\end{figure}
\begin{figure}[h]
\centering
\includegraphics[width=\textwidth]{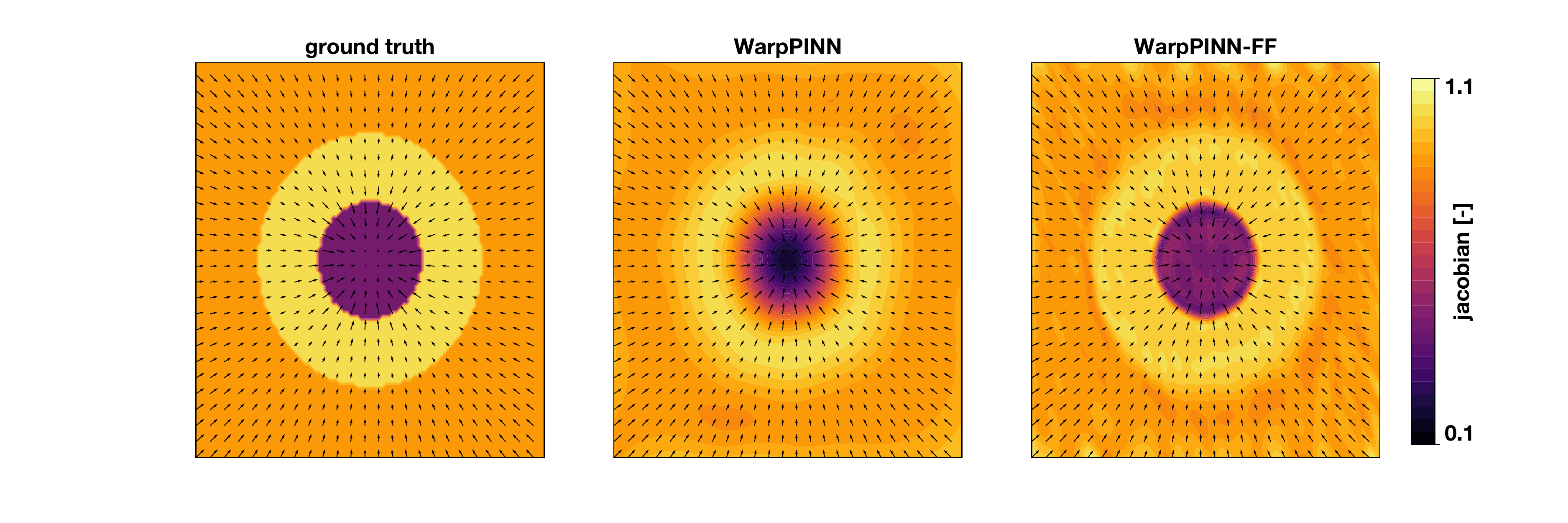}
\caption{Ground truth jacobian (left) versus predicted jacobians with WarpPINN (center) and WarpPINN-FF (right). Black arrows indicate the ground truth deformation field $\vec{u}$.}
\label{fig:jac}
\end{figure}

Template images warped with the predicted deformation field using WarpPINN and WarpPINN-FF are shown at the bottom left and right in figure \ref{fig:Reference and Target ring} respectively. By looking at the center of the images, it is easily seen that WarpPINN-FF has more capability to approximate the reference image. The mean squared error and SSIM is improved from $6.94\times10^{-4}$ and 0.984 respectively with WarpPINN to $1.84\times10^{-4}$ and 0.996 with WarpPINN-FF. For better visualization, profiles of the obtained deformation fields, strains and jacobian versus the ground truth are shown in figure \ref{fig:jac slice}, where the effect of using Fourier features is enhanced: even though both WarpPINN and WarpPINN-FF make a fair prediction of the displacement field $\vec{u}$, the first performs poorly when taking the derivative to compute the strains and the jacobian, being unable to capture the sharp features of the deformation field at the boundaries of the ring. Fourier features in turn are allowing our neural network to improve the prediction by capturing these discontinuities at the cost of introducing more oscillations. We highlight that, as commented before, it is necessary to only use $m=8$ features to make more accurate predictions. Also, the spectral bias for neural networks is shown in figure \ref{fig:jac}, where WarpPINN predicts a smooth Jacobian as opposed to WarpPINN-FF, which captures the sharp discontinuities on the boundary of the ring but also introduces some oscillations in the outer region.

\section{Registration of cine-MRI}

To validate our approach, we apply WarpPINN on a publicly available dataset through the Cardiac Atlas Project\footnote{\url{http://www.cardiacatlas.org/challenges/motion-tracking-challenge/}}. It corresponds to a motion tracking challenge proposed by the MICCAI on the workshop “Statistical Atlases and Computational Models of the Heart: Imaging and Modelling Challenges” (STACOM’11) held in 2011, where the goal is to accurately quantify the motion of the left ventricle during the cardiac cycle from different imaging modalities. The dataset consists of 15 healthy volunteers imaged with cine SSFP and tagged magnetic resonance, and 3D ultrasound. For each volunteer there are 12 manually tracked landmarks by two observers using tagged MRI, which are considered the ground truth to assess tracking methods. Also, a mesh representing the segmentation of the left ventricle at end diastole is provided, which we use to identify the incompressibility region in the domain.

A benchmarking framework for this challenge has been introduced in \cite{Benchmark} where the results over cine MR images of two other methods are presented, the Temporal Diffeomorphic Free Form Deformation (TDFFD) \cite{UPF} and ILogDemons \cite{INRIA}. Additionally, we consider CarMEN \cite{CarMEN}, another deep learning approach to cardiac motion, which uses a fully connected neural network. This will provide a direct comparison to evaluate the proposed method, where the main metrics to assess their accuracy are landmark tracking and strain curves.

\subsection{Extending WarpPINN to 3D+t}

Cine MR images correspond to a sequence of $N_f$ frames spanning the cardiac cycle, where each frame is a stack of 2D images. This motivates to include the time $t$ as an additional input for the neural network, thus, we are looking for the deformation field that for a given point $(X,Y,Z)$ in the reference frame and time $t$, outputs its new location $(X,Y,Z) + u(X,Y,Z,t; \vec{\theta})$ at time $t$. We set our reference image $R$ to be the first frame of the whole sequence at time $t=0$, corresponding to end diastole, which is when the blood volume in the ventricles is the largest. The time $t$ is considered to be in the interval $[0,1]$, such that the rest of the frames are taken as template images at equispaced different times, with the $N_f$-th frame corresponding to $t=1$. By doing so, the displacement field from the reference state $R$ to each of the following frames is learnt. Let $T_j$ be the $j$-th frame, and $t_j = j/(N_f-1)$ be the corresponding time, for $j\in \{0,\ldots,N_f-1\}$, then we consider the similarity metric between the reference image $R$ and $T_j$ for $p=1$:
\begin{equation}\label{eq:loss function p=1}
    \mathcal{L}_j(\vec{\theta}) = \frac{1}{N}\sum_i^{N} |R(\vec{X}^d_i) - T_j(\vec{\varphi}(\vec{X}^d_i, t_j;\vec{\theta}))|.
\end{equation}

To evaluate the neo-Hookean regularization term we consider
sampled points $\{(\vec{X}^{\rm inc}_l, t_l)\}_{l=1}^{N_{\rm inc}} \subset \Omega_{\rm inc}\times [0,1]$ and $\{(\vec{X}^{\rm bg}_k,t_k)\}_{k=1}^{N_{\rm bg}}\subset \Omega_{\rm bg}\times [0,1]$ as the collocation points at the incompressible and background regions respectively. The points $\{\vec{X}^{\rm inc}_l\}_{l=1}^{N_{\rm inc}}$ and  $\{\vec{X}^{\rm bg}_k\}_{k=1}^{N_{\rm bg}}$ are easily obtained as the challenge provides a segmentation of the left ventricle. The times $\{t_l\}_{l=1}^{N_{\rm inc}}$ and $\{t_k\}_{k=1}^{N_{\rm bg}}$ are randomly taken in $[0,1]$. The loss function is finally defined as
\begin{equation}\label{eq:loss function}
    \mathcal{L}(\vec{\theta}) = \sum_j^{N_f}\mathcal{L}_j(\vec{\theta}) + \mu \left(\frac{1}{N_{\rm inc}}\sum_{l}^{N_{\rm inc}} W(\vec{\varphi}(\vec{X}^{\rm inc}_l,t_l; \vec{\theta});\lambda_{\rm inc}) + \frac{1}{N_{\rm bg}}\sum_{k}^{N_{\rm bg}} W(\vec{\varphi}(\vec{X}^{\rm bg}_k,t_k; \vec{\theta});\lambda_{\rm bg})\right).
\end{equation}

The composition of $T_j$ with the deformation field is performed with trilinear interpolation. This is computationally expensive, making unsuitable the evaluation of $\mathcal{L}_j$ for all $j$ at each iteration. Instead, only one frame $T_j$ of the sequence is randomly taken per iteration as template image, the network is fed with the corresponding time $t_j$ and the parameters are updated to solve the registration between both images. On the other hand, the evaluation is made by mini-batches on the collocation points $(\vec{X}^{\rm inc}_l, t_l)$ and $(\vec{X}^{\rm bg}_k, t_k)$.


\subsection{Computation and visualization of strains}

Thanks to the architecture of the neural network, computing the derivatives of the deformation field $\vec{u}$ with respect to the spatial variables $\vec{X}$ is straightforward with automatic differentiation. The myocardial strain tensor is then computed at any point $\vec{X}$ and time $t$ as
\[\ten{E}(\vec{X},t) = \frac{1}{2}\left( \ten{C} - \ten{I}  \right) = \frac{1}{2}\left(\ten{F}^T\ten{F} - \ten{I}  \right). \]
Then, this tensor is projected along a set of directions $\vec{p}$, corresponding to the radial, longitudinal, and circumferential vectors of a local coordinate system, and related to the anatomy of the left ventricle. The longitudinal direction $\vec{l}$ was defined uniformly by drawing a line from the apex to the mitral valve. The provided mesh of the left ventricle is aligned along this direction, hence $\vec{l}=(0,0,1)$. To obtain the radial direction $\vec{r}$, first it is computed the normal vector $\vec{e}$ to each node of the mesh and then the vertical component is subtracted, then $\vec{r} = \vec{e} - (\vec{e}\cdot \vec{l})\vec{l}$. The circumferential direction $\vec{c}$ is directly computed as the cross product between the previous two vectors $\vec{c} = \vec{l}\times\vec{r}$. Finally, the strain tensor along this set of directions is computed as 
\[\ten{E}_{\vec{p}}(\vec{X},t) = \vec{p}^T\cdot\ten{E}(\vec{X},t)\cdot\vec{p}.\]

To visualize the strain curves the 17 American Heart Association (AHA) segments are used following the guidelines in \cite{Benchmark}: for each anatomical direction and at each time, strain values are averaged over several AHA segments. The left ventricle mesh provided by the challenge contains also the labels for each segment. In particular, we consider those segments corresponding to the septum and free wall. Finally, these averages are plotted against time. 

\subsection{Alternative methodologies for comparison.}



For comparison, we briefly introduce three different image registration algorithms that were applied over the same dataset. The Temporal Diffeomorphic Free Form Deformation was developed by researchers at the Universitat Pompeu Fabra (UPF) \cite{UPF}. In this approach a 3D+t velocity field is parametrized by B-Spline spatiotemporal kernels, the displacement field is then retrieved by forward Euler integration of this field. The mean squared error between the intensities of a chosen reference frame and the rest of them is used as for image similarity, and  quasi-incompressibility is imposed by adding a regularizer to the cost function that favours a zero-divergence velocity field. We refer to this method as UPF. The second method is called iLogDemons and was proposed by the Inria-Asclepios project \cite{INRIA}. It is a modification of the logDemons algorithm to integrate elasticity and incompressibility for cardiac motion tracking, where a non-linear registration method is performed through a diffeomorphic transformation parameterized by a stationary velocity field. Incompressibility is fully imposed via a divergence-free constraint. We refer to this method as INRIA. The third method taken into account is the recently published CarMEN \cite{CarMEN}, a convolutional neural network that takes two images as input, the reference and template images, and outputs the corresponding deformation field. The template image is deformed via trilinear interpolation, the L1-norm is used as similarity measure between both images, and a diffusion-regularizer based on the Jacobian of the deformation field. We refer to this method as CarMEN. Results for both, UPF and INRIA, are publicly available for the MICCAI workshop dataset, whilst for CarMEN, we ran the neural network on the dataset. 

\subsection{Results}

To assess the accuracy of a predicted deformation field, the dataset provides 12 ground truth landmarks per volunteer, one landmark per wall (anterior, lateral, posterior, septal) per ventricular level (basal, midventricular, apical). These landmarks were generated from 3DTAG datasets and then manually-tracked by two observers. As a post-process, these landmarks were converted to SSFP coordinates with DICOM header information. Due to temporal miss alignment between these two modalities, landmarks in SSFP coordinates can be compared at two times only: end systole and end diastole (final frame). Ground truth landmarks at the first frame are used as initialization points as they are in the reference frame, then the trained neural network is ran over these points to predict their position at every frame. Finally, the euclidian error between the obtained deformed landmarks and the two manually-tracked landmarks at the remaining two frames is measured and the errors obtained by WarpPINN and all the alternative methodologies previously introduced are compared. 

Additionally to landmark tracking, we also present the radial, longitudinal and circumferential strain curves output by our method. However, in this case, there is no ground truth to compare with, hence, we assess our results according to the literature and by comparing with UPF and INRIA results only. 

As with any neural network, we can overfit the data, which can lead to a low error in landmarks, but a largely irregular deformation field and non-realistic strain curves. In this sense, the regularizer $\mu$ plays an important role in the results, but its choice is not clear. In the rest of this section we consider the cases $\mu = 10^{-5}$ or $\mu=5\times 10^{-6}$, and $\sigma = 1$. Just as in the 2D case, in the forthcoming figures we shall refer to as WarpPINN-FF and WarpPINN to our method with and without Fourier features respectively.

We run all experiments on a cluster using one NVIDIA QUADRO RTX 8000 GPU card. Training the neural network took 70 minutes approximately without Fourier features and 140 minutes with Fourier features.

\begin{figure}[!t]
\centering
\includegraphics[width=1\textwidth]{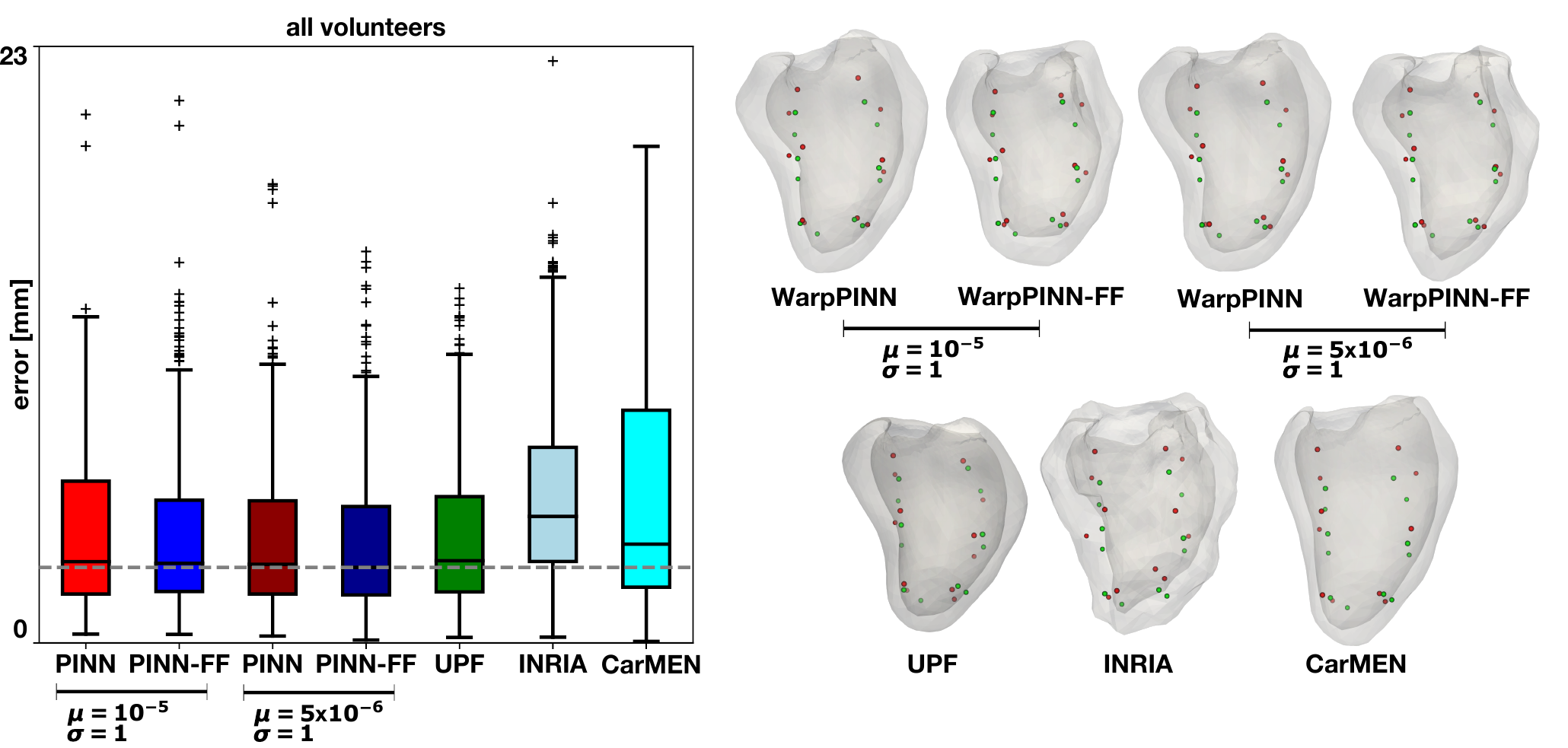}
    \caption{Left: box plots for landmark tracking errors. The dotted gray line indicates the lowest median, attained for WarpPINN-FF with $\mu=5\times 10^{-6}$ and $\sigma=1$. PINN refers to WarpPINN, PINN-FF refers to WarpPINN-FF. Right: Deformed meshes for volunteer 1 at end systole with predicted landmarks (in red) versus Ground Truth landmarks given by Observer 1 (in green) at End Sistole and for all methodologies.}
    \label{fig:lmks all volunteers}
\end{figure}

\subsubsection{Landmarks}

The distribution of the error over all the volunteers between manually-tracked and deformed landmarks by different methods is presented in box-plots in figure \ref{fig:lmks all volunteers}. The medians for UPF, INRIA and CarMEN are 3.17 mm, 4.87 mm, and 3.8 mm respectively. Using $\mu=10^{-5}$ and $\sigma = 1$ the medians for WarpPINN with and without Fourier features are 3.06 mm, and 3.13 mm respectively; using $\mu=5\times 10^{-6}$ and $\sigma = 1$ the errors for WarpPINN with and without Fourier features are 2.91 mm, and 3.02 mm. From here, it can be seen that decreasing the value of the regularizer $\mu$, implies a better performance for landmark tracking as larger deformations are being allowed. Also, for both regularizers, the performance of WarpPINN is comparable to that of UPF, with a slightly better median for both cases, with and without Fourier feature mappings. Outliers in the first two methods in figure \ref{fig:lmks all volunteers} correspond to two landmarks for volunteer v2 that are slightly deviated from the left ventricle mesh in the first frame, hence, these are in the background region where our algorithm allows larger deformations. Additionally, figure \ref{fig:lmks all volunteers} shows the predicted state of the left ventricle of volunteer v1 at end systole for all methods as well as the ground truth landmarks for observer 1 (in green) and predicted landmarks (in red). From this figure it can be appreciated how difficult is for all methods to accurately predict the longitudinal deformation particularly at the basal level, with Fourier feature mappings improving the accuracy of the deformation field in this region.

\subsubsection{Strain curves and Jacobian}

\begin{figure}[t]
\centering
\includegraphics[width=0.9\linewidth]{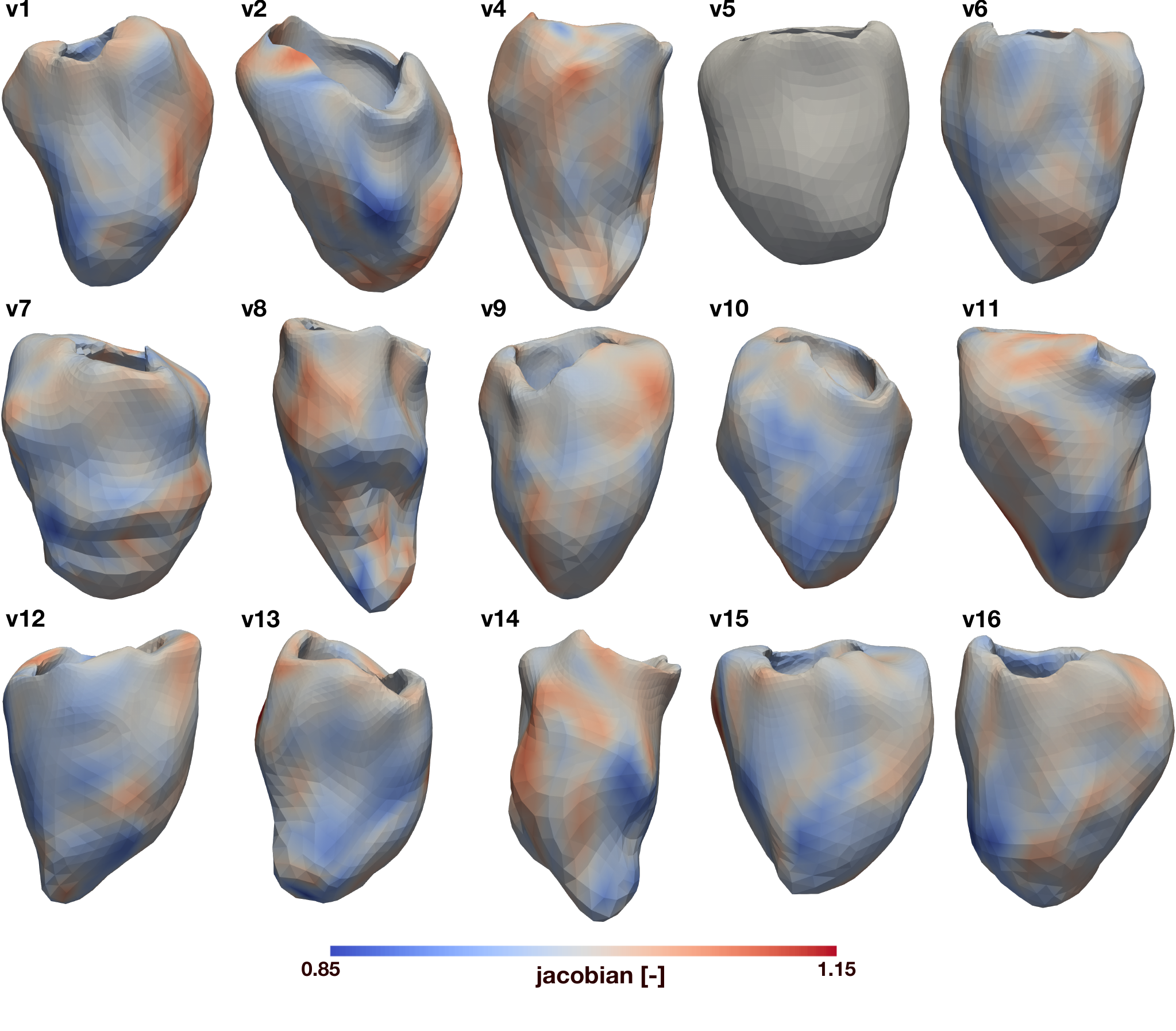}
\caption{Jacobian obtained by WarpPINN at end-systole with regularizer $\mu=10^{-5}$ for all volunteers.}
\label{fig:1e-05 jac es}
\end{figure}

\begin{figure}[t]
\centering
\includegraphics[width=0.9\linewidth]{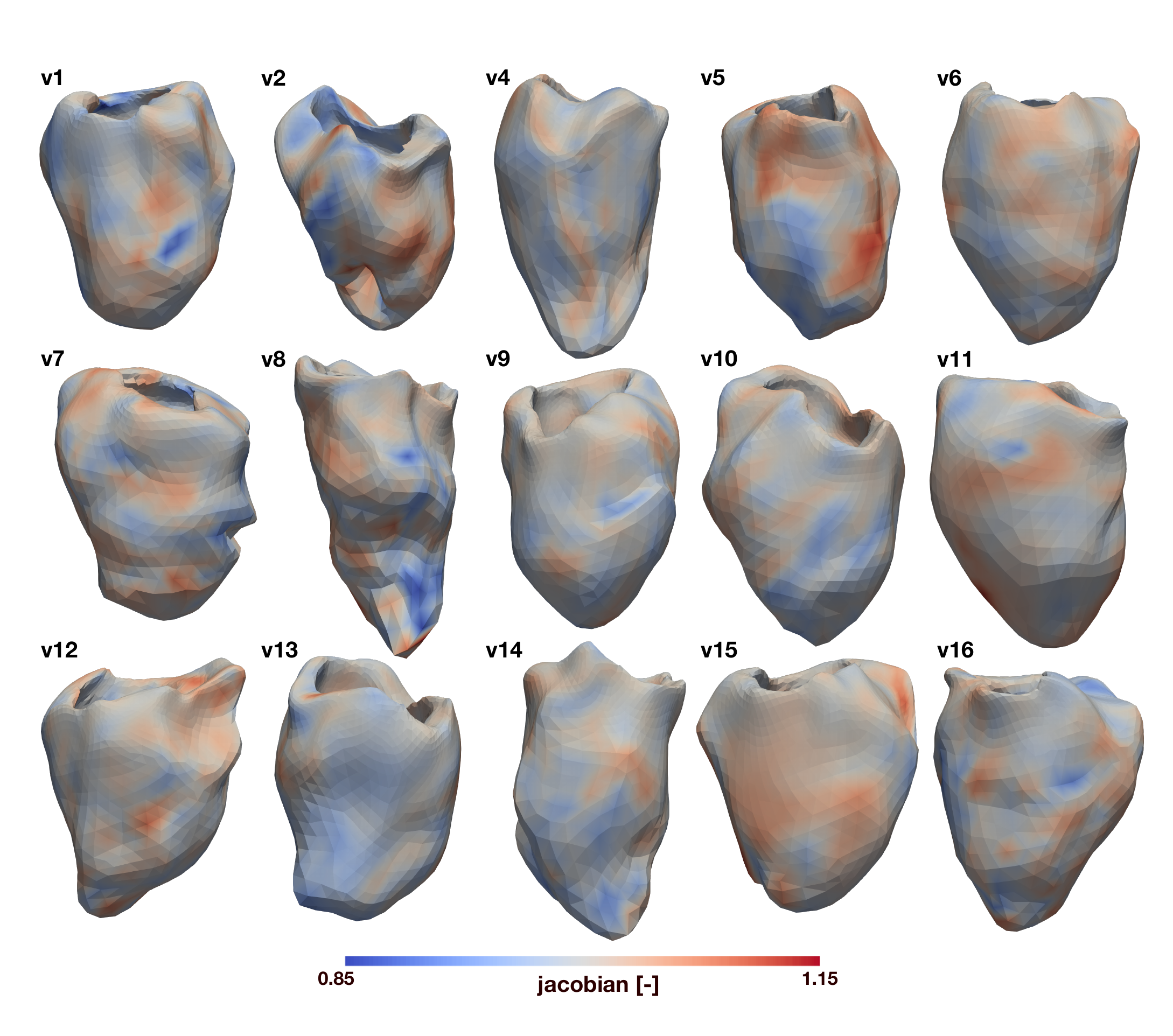}
\caption{Jacobian obtained by WarpPINN-FF at end-systole with regularizer $\mu=10^{-5}$ and $\sigma=1$ for all volunteers.}
\label{fig:1e-05 jac es ff}
\end{figure}

The Jacobian at end-systole for both WarpPINN and WarpPINN-FF using $\mu=10^{-5}$ and $\sigma=1$ are displayed on the deformed mesh for each volunteer in figures \ref{fig:1e-05 jac es} and \ref{fig:1e-05 jac es ff} respectively. For the Jacobian the colormap ranges from 0.85 to 1.15, with the neutral color (bone) representing regions where the Jacobian is equal to 1, blue colors representing negative values and red colors representing positive values. Figures \ref{fig:1e-05 jac es} and \ref{fig:1e-05 jac es ff} show that the values obtained for all the volunteers are all close to 1, being satisfied the nearly incompressibility imposed through the Neo-Hookean. In the case of volunteer v5, WarpPINN with $\mu = 10^{-5}$ is unable to perform registration properly, leading to a deformation field close to the identity, thus, the Jacobian is 1 almost everywhere. Using Fourier Feature mappings leads to larger deformations and volunteer v5 is warped, however, volunteers v2, v7, and v12 present unrealistic deformations. 

\begin{figure}[t]
\centering
\includegraphics[width=1\linewidth]{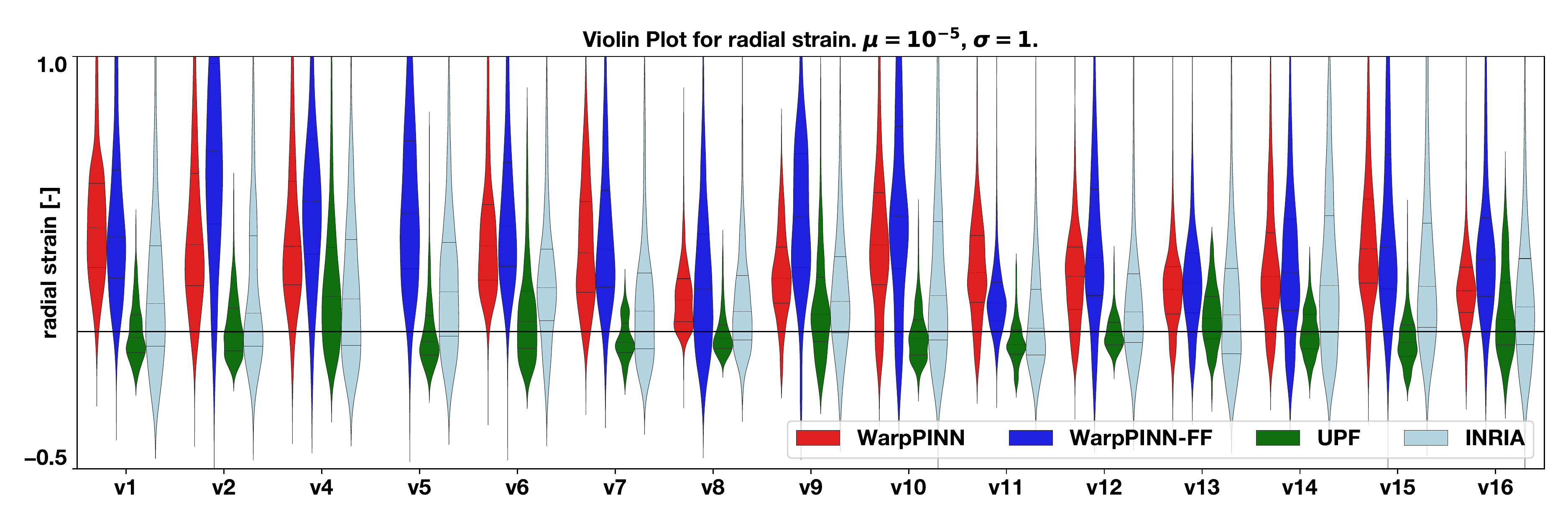}
\caption{Violin plots for radial strain values for all segments at end-systole for all volunteers and all methodologies. WarpPINN with $\mu=10^{-5}$, WarpPINN-FF with $\mu=10^{-5}$ and $\sigma = 1$}
\label{fig:violin plot}
\end{figure}

\begin{figure}[!htbp]
    \centering
    \hspace{-10mm}
    \includegraphics[scale = 0.84]{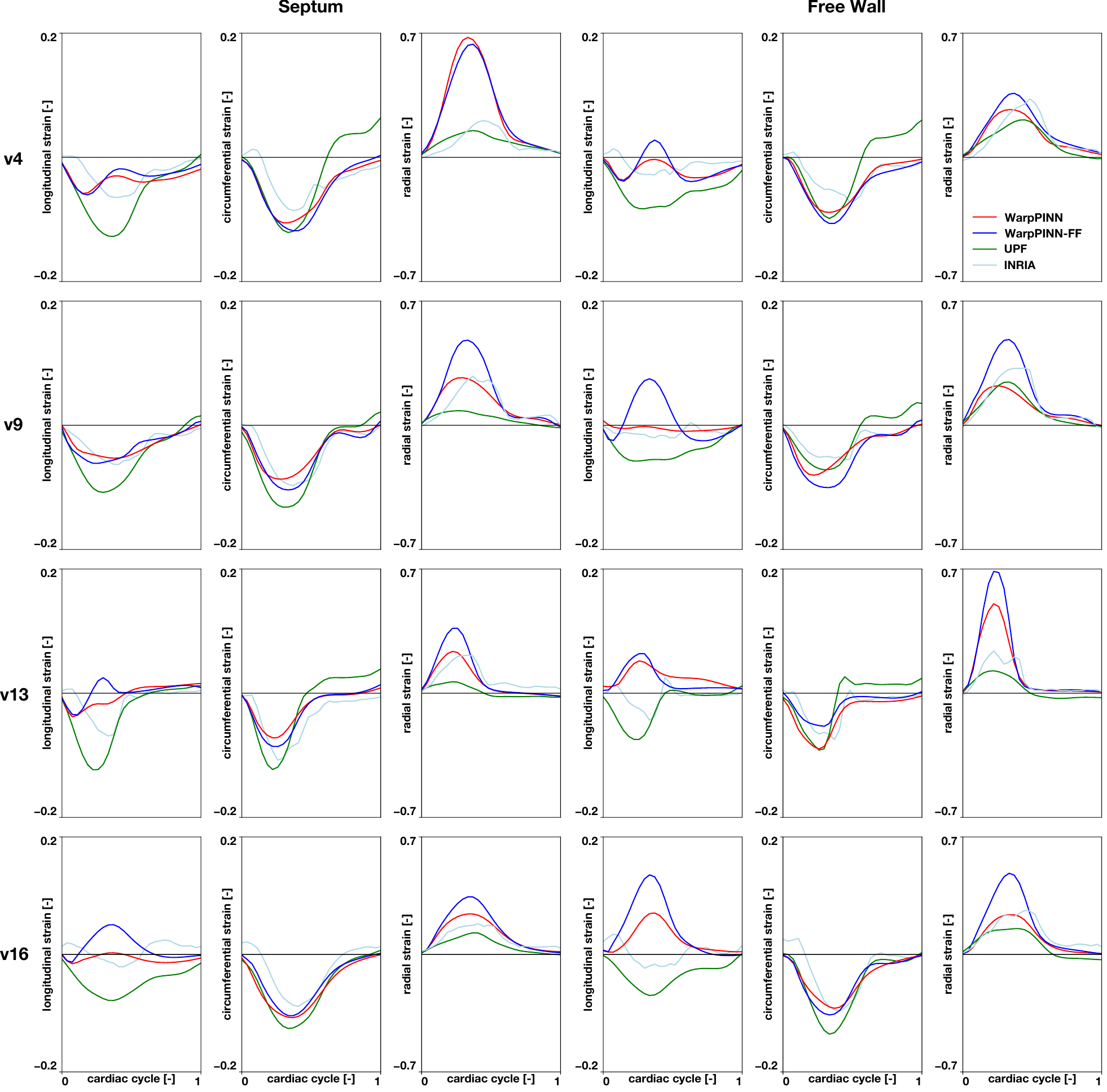}
    \caption{Strain curves. Volunteers by row: v4, v9, v13, and v16. By column: longitudinal, circumferential and radial strains in septum and free wall respectively. Red: WarpPINN with $\mu=10^{-5}$, blue: WarpPINN-FF with $\mu=10^{-5}$ and $\sigma = 1$, green: UPF, gray: INRIA.}
    \label{fig:strain curves}
\end{figure}

On the other hand, the distribution of radial strain in the whole ventricle at end-systole are depicted in figure \ref{fig:violin plot} through violin plots. There, we compare our method with of UPF and INRIA, and it is appreciated how our method predicts larger values for the radial strain. This is also confirmed in figure \ref{fig:strain curves} where the strain curves for 4 volunteers are depicted.


For strain quantification, we follow \cite{Benchmark}, where instead of showing the curves for every AHA segment, these values are averaged in two regions, septum (segments 1, 2, 3, 7, 8, 9, 13, 14) and free wall (segments 4, 5, 10, 11, 15, 16). Strain curves on these segments are depicted in figure \ref{fig:strain curves}. Some advantages of WarpPINN are the smoothness of these curves (as opposed to INRIA), the high peaks for the radial strain, close to physiological values of 45\% (as opposed to UPF), and the lack of drift effect at the end of the cardiac cycle. The smoothness of the curve is explained through the spectral bias in physics-informed neural networks: in the time variable we are not using Fourier feature mappings, thus low frequency behaviours are dominant in this variable. 
Finally, the lack of drift is explained through the loss function, where the reference image is always taken as the first frame. 

\section{Discussion}

Identifying heart failure from non-invasive measurements is a relevant task for the assessment of cardiac dysfunction. The evaluation of tissue strain seem to be a promising metric for cardiac diagnosis but obtaining them remains a challenging problem. On the other hand, cine SSFP MRI is, to the date, the gold-standard for heart imaging, making it desirable to get strain estimates from this data source. In this work, we use a fully connected neural network to perform image registration and apply it to cardiac strain assessment. WarpPINN can easily impose quasi-incompressibility through the determinant of the jacobian of the deformation field via automatic differentiation. Once the neural network has been trained, we learn a non-rigid deformation and then we can apply it to points in the myocardium to directly compute the strains, again, with automatic differentiation.  


Our method is able to accurately predict landmark deformation as can be seen in figure \ref{fig:lmks all volunteers}, where the median of the error distribution for WarpPINN in four different settings is better when compared to UPF, INRIA, and CarMEN, however the presence of outliers cannot be neglected. At the same time, it can be appreciated in figure \ref{fig:strain curves} the large values for the radial strain obtained by WarpPINN as opposed to those of UPF, and the smoothness of strain curves as opposed to, for instance, INRIA. Additionally, quasi-incompressibility is satisfied with values of the Jacobian in the ventricle close to 1 as shown in figures \ref{fig:1e-05 jac es} and \ref{fig:1e-05 jac es ff}. However, we also observe a common problem of image registration methods: the trade-off between fitting landmarks and obtaining realistic strains. In this sense, the choice of the regularizer $\mu$ plays an important role: the smaller it is, the larger the predicted deformations are, and then landmark tracking is accurately solved, however, this leads to large values of strains and more deviations from 1 of the Jacobian. The opposite occurs for larger values of $\mu$. We also note that even though we get the lowest median error for the landmark tracking with WarpPINN, it is still not perfect. Tracking landmarks manually is a non trivial task that can explain this issue: it is reported in \cite{Benchmark} that landmarks were acquired with tagged MRI and then registered to cine MRI leading to an inter-observer variability of 0.84mm between the two observers, meaning that, there is no total agreement even for manually tracked landmarks. 

We also emphasize the choice for the similarity metric used to compare the reference image with the warped template image. We found that the $L_2$ norm worked better for the synthetic example while the $L_1$ norm worked better for registration of cine SSFP MRI. An explanation for this behavior in cine SSFP MRI is that, during the cardiac cycle, there are some structures of the tissue appearing and disappearing in the sequence of images, such as the valves at the base. These are regarded as outliers that are largely penalized with the $L_2$ norm, which led to very large and unrealistic deformations in the longitudinal direction for the ventricle. The $L_1$ norm instead, penalizes uniformly these differences and then is not affected by outliers. This is a known issue in the literature \cite{nestares2000robust} and we will consider alternative loss functions especially designed for less sensitivity to outliers, such as the Tukey's biweight loss or the Hubber loss \cite{Tukey, Huber_1964}. 

Our approach can be easily generalized to other image registration settings and other physical constraints. Nonetheless, it presents some opportunities for improvement. For instance, we know that the predicted deformation mapping must be a diffeomorphism. This implies that the Jacobian is positive in the entire domain. Currently, we favor this behavior with our hyperelastic regulatization, but it is not guaranteed. We can overcome this issue by changing the architecture of the network into, for instance, invertible neural networks \cite{Ardizzone}. Another option is to predict the velocity field instead of the displacement field, as in \cite{INRIA, UPF}. However, both options will come with an additional cost during training since the computational graph becomes more complex and the gradient computation more expensive. Another limitation of our method that is present both in image registration methods and neural networks is the need of hyper-parameter tuning. The choice of the regularizer $\mu$ or the parameters in Fourier feature mappings is not clear and we may resort to automatized routines for choosing these parameters in the future. Nevertheless, it was seen that WarpPINN performed well on 15 different subjects for the same set of parameters. The exception for $\mu=10^{-5}$ was volunteer v5 where the predicted deformation was extremely close to the identity, leading to almost no deformation, however, a strong miss-alignment in slices in $z$ is appreciated in the MR images which makes it more difficult to perform image registration. Decreasing the value of the regularizer however allowed to avoid this behavior.

Our neural network learns the deformation field by taking $\vec{X}$ as input and then predicting its new position. This implies that WarpPINN needs to be trained for each case, as opposed to, for instance, CarMEN, which is trained to output the whole deformation field from the cine-MRI sequence as input in one shot. We show that we can obtain more accurate results with WarpPINN as each case is processed independently. In this direction, we plan to accelerate the training and prediction by exploring transfer learning \cite{lejeune2020exploring} and also to predict in one shot by encoding the input images while retaining the continuous representation of the displacements with architectures such as deep operator networks \cite{wang2021learning} or hypernetworks \cite{ha2016hypernetworks}. Overall, WarpPINN provides a solid first step towards accurate cardiac strain estimation from cine MRI and open doors to test this methodology in a wide range of image registration problems.

\section{Acknowledgments}

PA is supported by the scholarship from the EPSRC Centre for Doctoral Training in Statistical Applied Mathematics at Bath (SAMBa), under the project EP/S022945/1. HM acknowledges the support ANID - FONDECYT Postdoctorado \#3220266. This work was funded by ANID – Millennium Science Initiative Program – ICN2021\_004  to HM, SU and FSC, and NCN19\_161 to FSC. FSC also acknowledges the support of the project FONDECYT-Iniciaci\'on 11220816.



\newpage

\bibliographystyle{elsarticle-harv}
\bibliography{litra}

\appendix

\section{Construction of numerical example in 2D}

The construction of the deformation field used in section \ref{sec:2d} is as follows: a ring of thickness $R_2-R_1$ is deformed into another ring centered at the same point but with thickness $r_2-r_1$ such that the volume (area) of both rings remains the same, that is
\begin{equation}\label{eq:same area 1}
    R_2^2-R_1^2 = r_2^2-r_1^2.
\end{equation}

The proposed radial deformation field has the form
\begin{equation}\label{eq:deformation phi}
    \vec{\varphi}(X,Y) =  f\left(\sqrt{X^2+Y^2}\right)\begin{pmatrix}
X\\Y
\end{pmatrix},
\end{equation}
for some function $f$ to be determined. If we want $\vec{\varphi}$ to be incompressible in the ring $R_1^2<X^2+Y^2<R_2^2$ we have to impose $\det(J\vec{\varphi}(X,Y))=1$ there. Seeing $f$ as a function of $R=\sqrt{X^2+Y^2}$, then this condition implies that $f$ satisfies the following ODE:
\begin{equation}\label{eq:ode}
    f'(R) + \dfrac{f(R)}{R} = \dfrac{1}{Rf(R)}.
\end{equation} 
Moreover, we want that if a point $(X,Y)$ is on the inner ring, that is, $X^2+Y^2 = R_1^2$, then, $\vec{\varphi}$ has to satisfy the next equality:
\[\vec{\varphi}(X,Y) =  \dfrac{r_1}{R_1}\begin{pmatrix}X \\ Y\end{pmatrix}.\]
This vector has length $r_1$ in the direction of the vector $(X,Y)$ which is normalized by $R_1$. Hence, we impose the next boundary condition 
\begin{equation}\label{eq:ode bc}
    f(R_1) = \dfrac{r_1}{R_1}.
\end{equation}
The solution of the ODE \eqref{eq:ode} with condition \eqref{eq:ode bc} is
\begin{equation}\label{eq:ode solution}
    f(R) = \dfrac{1}{R}\sqrt{R^2-R_1^2+r_1^2}, \quad R_1<R<R_2.
\end{equation}
Finally, the deformation $\vec{\varphi}$ in \eqref{eq:deformation phi} is constructed from this radial function $f$:
\[f(R)=\left\{\begin{array}{ccl} 
\dfrac{r_1}{R_1} & \text{if} & R<R_1\\
\dfrac{1}{R}\sqrt{R^2-R_1^2+r_1^2}& \text{if} & R_1\leq R \leq R_2\\
\dfrac{r_2}{R_2} & \text{if} & R>R_2,\\
\end{array}\right.\]
where the values of $f$ outside the ring makes it a continuous function with discontinuities of the derivative on $R=R_1$ and $R=R_2$.


\end{document}